\def\aa{A\&A}               %Astronomy & Astrophysics%
\def\anj{AJ}                %Astronomical Journal%
\def\apj{ApJ}               %Astrophysical Journal%
\def\apjs{ApJS}             %Astrophysical Journal Supplements
\def\baas{BAAS}             %Bulletin of the American A.S.%
\def\mn{MNRAS}              %Monthly Notices of the Royal...%
\def\mc{\multicolumn{2}{c}{--}}

\documentstyle{aa}

\begin{document}

\title{The inner kiloparsec of the jet in 3C\,264}
\subtitle{}

\author{L. Lara\inst{1,2} \and 
G. Giovannini\inst{3,4} \and
W.D. Cotton\inst{5}     \and
L. Feretti\inst{3}      \and
T. Venturi\inst{3}}

\offprints{L. Lara \email{lucas@ugr.es}}

\institute{Dpto. F\'{\i}sica Te\'orica y del Cosmos, Universidad de
Granada, Avda. Fuentenueva s/n, 18071 Granada (Spain)
\and Instituto de Astrof\'{\i}sica de Andaluc\'{\i}a (CSIC),
Apdo. 3004, 18080 Granada (Spain)
\and
Istituto di Radioastronomia (CNR), via P. Gobetti 101, 40129 Bologna (Italy)
\and
Dipartimento di Astronomia, Universit\'a di Bologna, via Ranzani 1,
40127 Bologna (Italy)
\and 
National Radio Astronomy Observatory, 520 Edgemont Road, Charlottesville, 
VA 22903-2475 (USA)
} 
\date{Received / Accepted}

\authorrunning{Lara et al.}
\titlerunning{The inner kiloparsec of the jet in 3C\,264}

\abstract{We present new multi-frequency EVN, MERLIN and VLA observations 
of the radio source 3C\,264, sensitive to linear scales ranging from
the parsec to several kiloparsecs. The observations confirm the
existence of regions with different properties in the first kiloparsec
of the jet.  The most remarkable feature is the transition between a
well collimated narrow jet at distances from the core below 80 pc, to
a conical-shaped wide jet, with an opening angle of
$20^{\circ}$. Another change of properties, consisting of an apparent
deflection of the jet ridge line and a diminution of the surface
brightness, occurs at a distance of $\sim 300$ pc from the core,
coincident with the radius of a ring observed at optical
wavelengths. Our observations add new pieces of information on the
spectrum of the radio-optical jet of 3C\,264, with
results consistent with a synchrotron emission mechanism and a
spectrum break frequency in the infrared.  Brightness profiles taken
perpendicularly to the jet of 3C\,264 are consistent with a spine
brightened jet at distances below 100 pc from the core, and an
edge-brightened jet beyond, which can be interpreted as evidence of a
transverse jet velocity structure. Our observations do not allow us to 
distinguish between the
presence of a face--on dust and gas disk at the center of the host
galaxy of 3C\,264, or rather an evacuated bubble. However, the
properties of the jet structure, the changes in the polarization
angle, and the plausible jet orientation can be naturally brought into
agreement in the bubble scenario.
}

\maketitle

\section{Introduction}

The radio source \object{3C\,264} (B1142+198) is a low luminosity
radio galaxy optically identified with NGC\,3862, an elliptical galaxy
in the cluster Abell 1367 with $m_v = 13.67$ and a redshift 
$z=0.0217$ (Smith et al. \cite{smith}). This radio galaxy has been 
observed at many different wavelengths from the radio to 
the X-ray, allowing a good determination of its nuclear 
spectral energy distribution (Capetti et al. \cite{capetti2}).
3C\,264 has a Fanaroff-Riley
type I radio structure (Fanaroff \& Riley \cite{fanaroff}) with a
head-tailed morphology at kiloparsec scales\footnote{We assume
throughout a Hubble 
constant of $H_0=72$ km s$^{-1}$ Mpc$^{-1}$ and a deceleration
parameter of $q_0=-0.55$, so that $1''$ corresponds to 412 pc}, a
prominent core and a wiggling jet 
extending toward the northeast that ends in a blob of emission at
$28''$ (11.5 kpc) from the core (Gavazzi, Perola \& Jaffe
\cite{gavazzi}; Bridle \& Vall\'ee \cite{bridle}; Baum et
al. \cite{baum1}; Lara et al. \cite{lara97}). There is evidence of
counterjet emission in the southwest direction from the core (Lara et
al. \cite{lara97}). Both the jet and counterjet are embedded in a vast
and diffuse region of low surface brightness emission which seems to
be dragged 
toward the north, possibly revealing the existence of a high density
intra-cluster medium.

Simultaneous EVN\footnote{EVN: European VLBI Network} and
MERLIN\footnote{MERLIN: Multi-Element Radio Linked Interferometer
Network} observations at 5.0 GHz by Lara et al. (\cite{lara99}) showed
for the first time the detailed structure of 3C\,264 at sub-kiloparsec
scales. It consists of a one-sided jet with evident variations in its
morphological properties with distance: {\em i)} the strong core and
innermost jet (0 - 10 pc); {\em ii)} a well-collimated and narrow
region (10-100 pc); {\em iii)} a region with strong widening, kinks
and filaments (100-300 pc); {\em iv)} a faint and narrow region after
a jet deflection (300 - 400 pc from the core).

Besides this rich radio structure, another peculiarity of 3C\,264 is
the existence of an optical jet coincident with the radio jet, most
possibly of synchrotron nature (Lara et al. \cite{lara99}) and
extending up to $2''$ (0.82 kpc) (Crane et al. \cite{crane}; Baum et
al. \cite{baum2}). There is also evidence of a ``ring'' around the
host galaxy core with a radius between $0''.75$ and $1''$ (300 - 400
pc) (Baum et al. \cite{baum2}).

We present in this paper new observations of the radio galaxy 3C\,264,
made with MERLIN and the EVN at 1.6 GHz, and VLA\footnote{VLA: Very
Large Array} observations at 1.6, 5.0, 22.5 and 43.3 GHz. The new data
allow a direct comparison with previous sub-kiloparsec scale
observations at 5.0 GHz (Lara et al. \cite{lara99}) and 
shed light on the properties of the inner kiloparsec of the jet in
3C\,264, a length scale which is crucial to the understanding of the
FR I vs. FR II 
dichotomy of radio galaxies.

\section{Observations and data reduction}

We observed 3C\,264 simultaneously with the EVN and MERLIN arrays
at 1.6 GHz on June 3rd 1997. Eleven EVN antennas participated in the
Very Long Baseline Interferometry (VLBI) observations forming a
sensitive interferometric array with minimum and maximum baselines of
0.77 M$\lambda$ and 46 M$\lambda$, respectively (Table~\ref{obs}). All
the antennas recorded left circular polarization, with a synthesized
bandwidth of 28 MHz. The strong compact sources 4C\,39.25 and 1144+402 were
observed to serve as fringe finders during the correlation of the
data.  The VLBI data were correlated in absentia by the staff of the
MkIIIA VLBI correlator at the Max-Planck-Institut f\"ur
Radioastronomie in Bonn (Germany). We used the NRAO
AIPS\footnote{AIPS: Astronomical Image Processing System, developed
and maintained by the NRAO} package to correct for instrumental phase
and delay offsets between the separate baseband converters in each
antenna, and to determine the antenna-based fringe correction. The
visibility amplitudes were calibrated using the system temperatures
and gain information provided for each telescope. Data imaging was
performed with the Difmap package (Shepherd et al. \cite{shepherd}). A
summary of these and other observations presented in this paper is
displayed in Table~\ref{obs}. 

\begin{table}[]
\caption[]{Summary of the observations}
\label{obs}
\begin{tabular}{lllll}
\hline
Array        & Frequency &  Bandwidth & Time   & Date    \\
             & (GHz)     &  (MHz)     & (min)  & (yr)    \\ \hline
EVN$^{a}$    & 1.6       &   28       & 720    & 1997.42 \\
MERLIN$^{b}$ & 1.6       &   16       & 720    & 1997.42 \\
VLA-A        & 1.6       &   50       &  30    & 1999.68 \\
VLA-A        & 5.0       &  100       &  30    & 1999.68 \\
VLA-A        & 22.5      &  100       & 180    & 1999.68 \\
VLA-A        & 43.3      &  100       & 180    & 1999.68 \\
\hline
\multicolumn{5}{l}{\parbox{8cm}{\footnotesize{$^a$ EVN array (diameter
in parenthesis): Shanghai (25m) and Urumqi (25m) in China, 
Simeiz (22m) in Ukraine, Torun (32m) in Poland, Medicina (32m) and
Noto (32m) in Italy, Effelsberg (100m) in Germany, Onsala (25m) in
Sweden, Westerbork (25m) in The Netherlands, and Cambridge (32m) and
Lovell-Jodrell (76m) in the U.K.}}}\\
\multicolumn{5}{l}{\parbox{8cm}{\footnotesize{$^b$ MERLIN array (diameter in
parenthesis): Defford (25m), Cambridge
(32m), Knockin (25m), Darnhall (25m), Jodrell-MK2 (25m),
Jodrell-Lovell (76m) and Tabley (25m)}}}
\end{tabular}
\end{table}

The MERLIN array consisted of seven antennas, all located in England
(U.K.) (Table~\ref{obs}). The maximum and minimum
baselines of the array were 1.2 M$\lambda$ and 37 k$\lambda$,
respectively. All telescopes recorded right and left circular
polarizations, excepting Cambridge which was participating also as
part of the EVN array and could not record both hands due to bandwidth
limitations in the radio link transporting the signal from each
MERLIN antenna to the correlator in Jodrell Bank. The data were edited and
amplitude calibrated at Jodrell Bank using standard procedures
based on the OLAF package (Muxlow et al. \cite{muxlow}). Flux density
calibration of the MERLIN data was performed by comparison of
\object{1144+402} and \object{OQ\,208} with the primary flux
calibrator \object{3C\,286}. The pipeline developed at Jodrell Bank
was used as a first approach for the phase calibration. The data were
then exported to the Difmap package, where we performed several cycles
of self-calibration.

\begin{figure}[]
\vspace{13.5cm}
\includegraphics{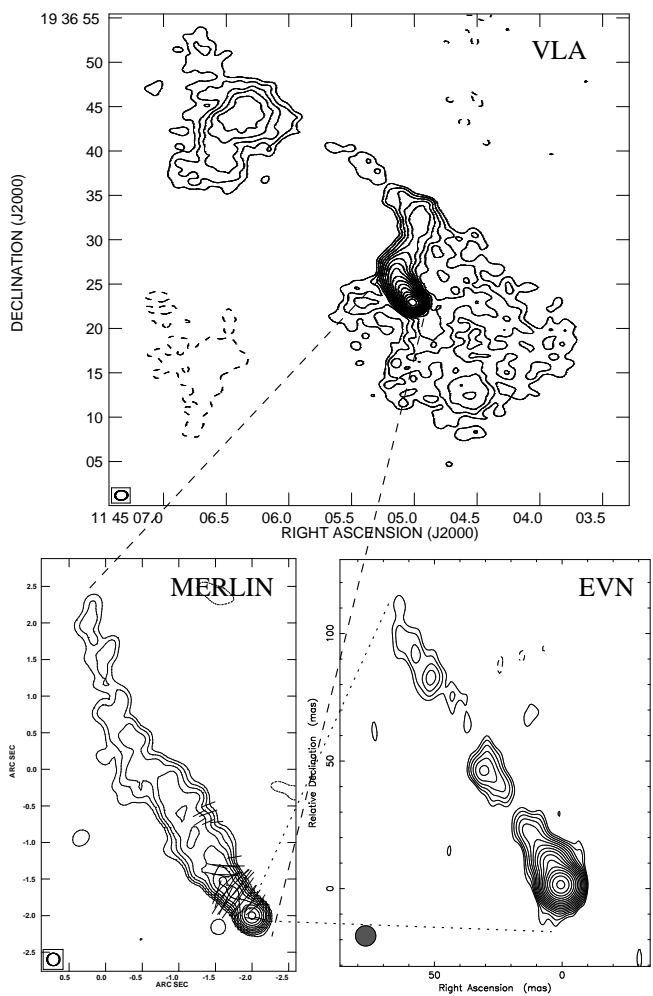}
%%\rule{0.4pt}{4cm}% line thickness, height of picture
\caption{Composition of VLA, MERLIN and EVN maps of 3C\,264 at 1.6 GHz. 
Contours are spaced by factors of
$\sqrt{2}$ in brightness, except in the MERLIN map where contours are
spaced by factors of 2. In this map we display vectors representing
the polarization position angle (E-vector), with length proportional
to the polarized flux ($1''\equiv 6.25$ mJy/beam). 
Dashed lines help to identify equivalent
regions in the different frames. For each map we list the Gaussian beam size 
used for convolution (in arcseconds), the first contour
level (mJy beam$^{-1}$) and the peak of brightness (Jy
beam$^{-1}$). {\bf VLA:} beam=$1''.39\times 1''.07$,
P.A. $-86.6^{\circ}$; 1st cntr = 0.85; peak = 0.278 {\bf MERLIN:}
beam=$0''.17\times 0''.17$; 1st cntr = 0.34; peak = 0.225 {\bf EVN:}
beam=$0''.008\times 0''.008$, ; 1st cntr = 0.64; peak = 0.127} 
\label{allmaps}
\end{figure}

The independently self-calibrated data from the EVN and MERLIN were
combined using the AIPS package. The different calibration procedures
applied to each data set resulted in a slight ($\sim 10\%$) misalignment in
the flux density scales of each interferometer. The antennas in
Cambridge and Jodrell Bank (Lovell) participated simultaneously in the
EVN and the MERLIN observations, defining a common baseline for the
two arrays which we used to match the flux density
scales. Since MERLIN is more sensitive to extended radio emission
which is resolved by the EVN and it makes use of primary flux density
calibrators during the calibration process, we adjusted the EVN data
to match the MERLIN flux density scale. 

3C\,264 was also observed at 1.6, 5.0, 22.5 and 43.3 GHz with the VLA
in its most extended configuration (VLA-A) on September 6th 1999. The
observing bandwidth was 50 MHz at 1.6 GHz and 100 MHz at the other
three frequencies. The whole VLA (27 antennas) participated in the
1.6 and 5.0 GHz observations. Only 13 antennas had 43 GHz receivers at
the time of the observations, so the array was split in two, with
these 13 antennas observing at 43 GHz and the remaining 14 observing
simultaneously at 22 GHz. At 1.6 and 5.0 GHz, the radio sources
\object{3C\,147} and \object{3C\,138} were used as primary
calibrators, and the nearby radio source \object{J1158+248} as
interferometric phase calibrator. At 22.5 and 43.3 GHz, 3C\,286 was
used as primary calibrator, and we used the fast switching mode during
the observations to reduce tropospheric phase fluctuations at 43
GHz. The nearby radio source \object{J1150+243} was used as a phase
calibrator. Moreover, the radio source
\object{J1159+292} was observed regularly at 8.4 GHz during the
observations to improve the pointing accuracy of the
interferometer at 22 and 43 GHz. The processes of self-calibration and
imaging of the data were carried out with the NRAO AIPS package and/or
Difmap following standard procedures.

\section{Results}

\subsection{Results at 1.6 GHz}

\begin{figure}[]
\vspace{12cm}
\includegraphics{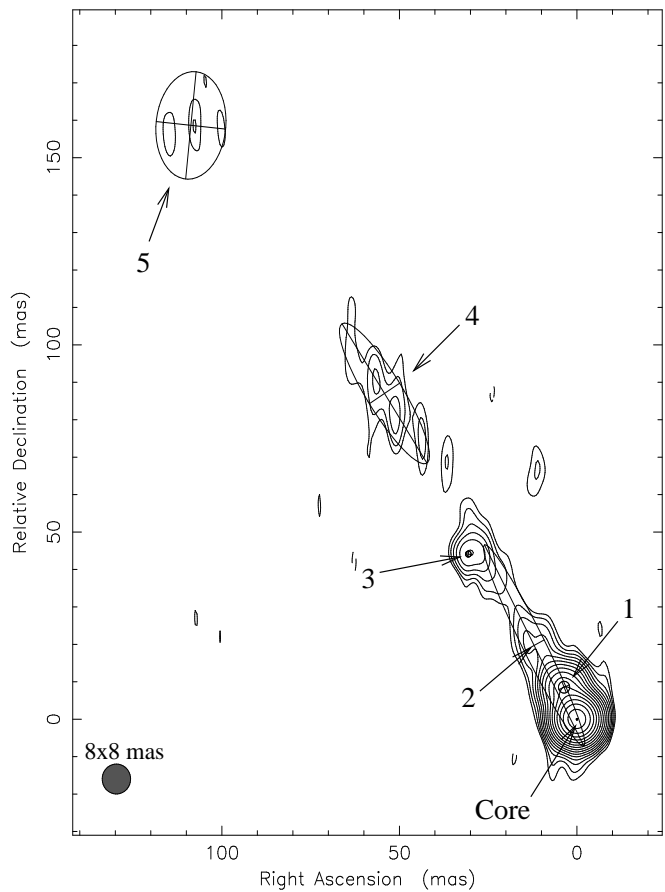} 
%%\rule{0.4pt}{4cm}% line thickness, height of picture
\caption{EVN map of 3C\,264 made at 1.6 GHz on 1997.42, with superimposed 
elliptical Gaussian components. Contours are spaced by factors of $\sqrt{2}$ 
in brightness. The circular Gaussian beam used for convolution is 8 mas 
$\times$ 8 mas. The first contour level is 0.77 mJy\,beam$^{-1}$ and the 
peak of brightness 0.128 Jy\,beam$^{-1}$.}
\label{modelfit}
\end{figure}

Images of 3C\,264 at 1.6 GHz made with the VLA, MERLIN and the EVN,
illustrating the very different linear scales sampled by each
interferometer separately, are displayed in Fig.~\ref{allmaps}. The
VLA map shows a strong core and a wiggling jet directed toward the
northeast. It confirms the existence of a weak counterjet in southwest
direction.  Moreover, it shows also the halo-like diffuse and extended
emission which surrounds both, the jet and counterjet. The map
obtained with MERLIN shows the jet of 3C\,264 up to $4''.5$ from the
core, where it starts its curvature toward the northwest as observed
with the VLA. The jet is broad and well resolved transversally, with a
width at zero intensity reaching $0''.8$ ($\sim 330$ pc)
at about $1''.2$ from the core. MERLIN detects significant polarized
flux in the jet up to a distance of $0''.8$ from the core with a mean
degree of polarization of $6.5\%$. The E-vector is oriented
perpendicularly to the jet ridge line (position angle, P.\,A. $\sim
-40^{\circ}$), with a rotation to P.\,A. = $-100^{\circ}$ at $0''.8$
from the core. The EVN map shows a core-jet parsec scale structure,
starting at P.\,A.$= +24^{\circ}$, slightly different from that
observed at larger scales (P.\,A.$=+35^{\circ}$). Several blobs of
emission can be distinguished in the parsec-scale jet, the most
remarkable one at 53.7 mas from the core and with a flux density of 4
mJy (labeled 3 in Fig.~\ref{modelfit}). 

\begin{figure*}[ht]
\vspace{8cm}
\includegraphics{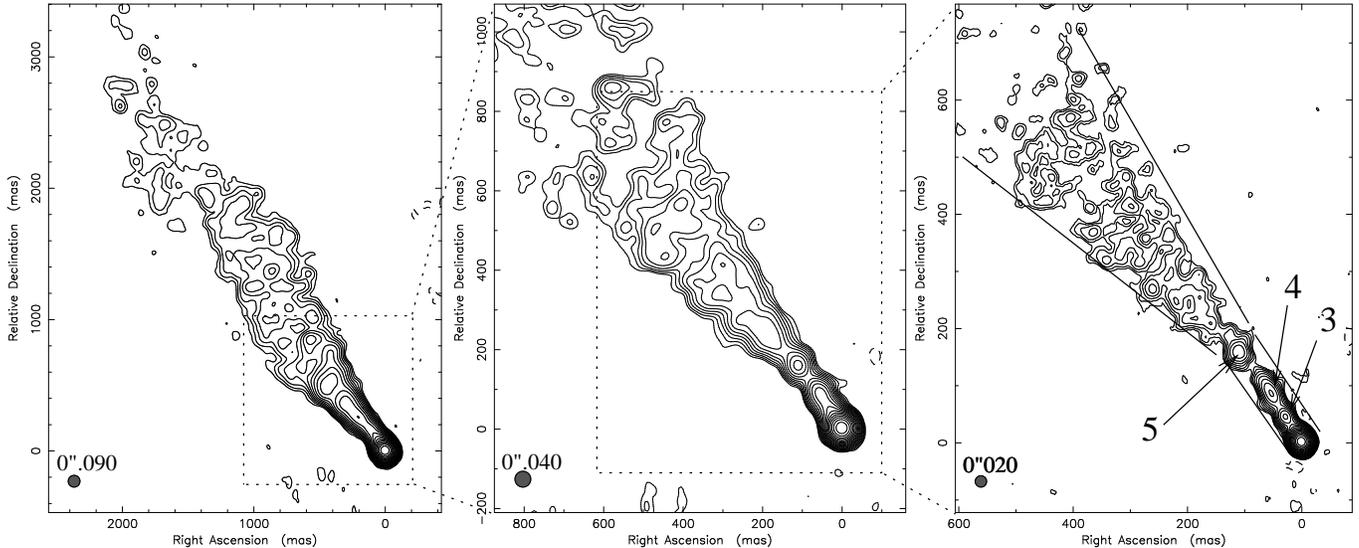}
%%\rule{0.4pt}{4cm}% line thickness, height of picture
\caption{MERLIN\,+\,EVN maps of 3C\,264 at 1.6 GHz. The three maps
have been obtained from the same data set, applying different Gaussian 
tapers to the visibility data. Circular Gaussian beams (bottom left of each 
frame) have been used for convolution. Contours are spaced by factors of
$\sqrt{2}$ in brightness. Dotted boxes help to identify equivalent
regions in the different frames.  For reference purposes, we label in
the map on the right the components ``3'', ``4'' and ``5'', detected
with the EVN 
(Fig.~\ref{modelfit}). The continuous lines on this map help to
identify the jet regions discussed in Section~\ref{long}. For
each map we list the first contour level 
(mJy\,beam$^{-1}$) and the peak of brightness (Jy\,beam$^{-1}$). {\bf
Left:} 1st cntr. = 0.42; peak = 0.209 {\bf Center:} 1st cntr. = 0.39;
peak = 0.195 {\bf Right:} 1st cntr. = 0.36; peak = 0.180.}
\label{combined}
\end{figure*}

In order to obtain a quantitative description of the parsec scale
structure of 3C\,264 which could be readily compared with future VLBI
observations, we have fitted simple elliptical Gaussian components to
the VLBI visibility data using a least square algorithm within Difmap.
Six components were required to satisfactorily reproduce the data. The
estimated parameters for each component are given in
Table~\ref{fit}. In Fig.~\ref{modelfit} we show the 
EVN map obtained after several iterations of model-fit and
self-calibration, with the model in Table~\ref{fit} superimposed.

\begin{table}[b]
\caption[]{VLBI components of 3C\,264$^{a}$}
\label{fit}
\begin{tabular}{c r@{.}l r@{.}l r@{.}l r@{.}l r@{.}l r@{.}l}
\hline
Comp.     & \multicolumn{2}{c}{S}   & \multicolumn{2}{c}{D}   &  \multicolumn{2}{c}{P.A.} &  \multicolumn{2}{c}{L}    & \multicolumn{2}{c}{r} & \multicolumn{2}{c}{$\Phi$} \\
          & \multicolumn{2}{c}{(mJy)} & \multicolumn{2}{c}{(mas)} & \multicolumn{2}{c}{(deg)} & \multicolumn{2}{c}{(mas)} & \multicolumn{2}{c}{}  & \multicolumn{2}{c}{(deg)}  \\
\hline
Core & 126&2  &   0&0  &   0&0 &  0&38  &  1&00 & \mc   \\
1    &  23&3  &   9&29 &  23&0 &  3&16  &  1&00 & \mc   \\
2    &  18&6  &  23&12 &  31&4 & 60&55  &  0&10 &  27&5 \\
3    &   4&0  &  53&69 &  34&7 &  1&59  &  1&00 & \mc   \\
4    &  11&9  & 102&53 &  31&9 & 44&22  &  0&23 &  32&7 \\
5    &   7&3  & 192&31 &  34&4 & 28&81  &  0&68 &$-$5&9 \\
\hline
\multicolumn{13}{l}{\parbox{8cm}{\footnotesize{$^a$ Symbols used are:
(S) flux density; (D) the angular distance from the core component;
(P.A.) the position angle with respect to the core; (L) the length of
the major axis; (r) the ratio between the major and minor axis;
($\Phi$) the orientation of the major axis defined in the same sense
as the P.\,A.}}} 
\end{tabular}
\end{table}

The advantage of combining two interferometers like MERLIN and the EVN
resides in that it is possible to obtain structure information at
angular resolutions intermediate to those provided by each instrument alone. In
Fig.~\ref{combined} we present three selected maps of 3C\,264 obtained
from the combined MERLIN\,+\,EVN data set. Sensitivity to emission at
different scales is obtained by applying the appropriate weighting
schemes to the combined visibility data during the mapping process,
this allows the mapping of the radio structure of 3C\,264 with angular
resolutions spanning the range 6 mas to 180 mas. The maps in
Fig.~\ref{combined} show the detailed structure of the jet of 3C\,264
in its first kiloparsec length.

\subsection{Results at 5.0 GHz}

\begin{figure}[]
\vspace{8cm}
\includegraphics{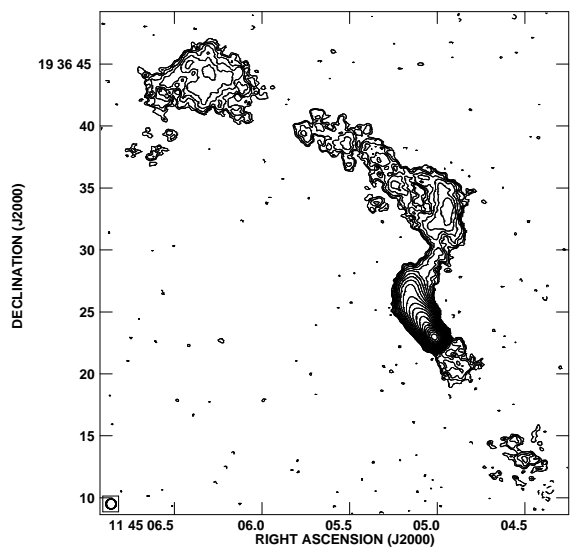}
%%\rule{0.4pt}{4cm}% line thickness, height of picture
\caption{VLA map of 3C\,264 made at 5.0 GHz on 1999.68. Contours are 
spaced by factors of $\sqrt{2}$ in brightness. The Gaussian beam used
for convolution is $0''.8 \times 0''.8$. The first contour level is 
0.13 mJy\,beam$^{-1}$
and the peak of brightness 0.236 Jy\,beam$^{-1}$.}
\label{bandac}
\end{figure}

Fig.~\ref{bandac} displays the VLA-A image of 3C\,264 made at 5 GHz,
with an angular resolution of $0''.8$. We can neatly follow the
curvature in the kiloparsec scale jet up to a distance of
17$''$ from the core. We detect the blob of emission at $28''$ from
the core in northeast direction, although at this angular resolution
we find no hints of the low level extended emission. Thanks
to the high sensitivity of the observations, we also detect radio
emission on the counterjet side, up to a distance of $5''$ from the
core.  The structure is faint and resolved transversally. We assume
that this feature contains the true counterjet. The fact that its
shape is not as collimated as expected in a jet is consistent with the
absence of a clear counterjet also in the VLA image at 1.6 GHz (upper
panel in Fig.~\ref{allmaps}). This might imply that the counterjet has
a large opening angle.\\

In Fig.~\ref{bandac_polz} we display the total intensity at 5 GHz with
vectors representing the polarized flux density (length) and
position angle (orientation). The angular resolution of this image is
$0''.6$. Most of the polarized emission comes from the unresolved
inner jet, where the mean degree of polarization is about 7\%. The
position angle in this region is different from that 
observed at 1.6 GHz, possibly due to Faraday rotation. At
larger distances, beyond 1'' from the core, the jet is well resolved
and the image shows that the
polarized emission is produced mostly at the jet edges. The degree of
polarization here is higher, reaching 15\%.

\subsection{Results at 22.5 and 43.3 GHz}

In Fig.~\ref{bandakq} we present the VLA-A images of 3C\,264 made at
22.5 and 43.3 GHz. The region shown corresponds approximately to the 
the dotted box in Fig.~\ref{combined}-left. We detect
the jet up to $0''.8$ and $0''.7$ from the core at 22.5 and 43.3 GHz,
respectively, and nothing beyond.  At 22.5 GHz we find polarized
emission in the jet, with a mean degree of polarization of 22\%. The
E-vectors are basically perpendicular to the jet ridge-line, except at
the observed jet extreme where there is an apparent change in
P.\,A. We also note the detection of polarized flux in the radio core,
with a degree of polarization slightly above the 1\% level. 
The polarization P.A. in the jet is not too different
from that found at 1.6 GHz, whereas the polarization percentage 
is about 3 times higher. 
This can be interpreted as due to Faraday rotation and implies 
that some P.A. rotation is actually present at 1.6 GHz, but it is not
apparent because of the $2\pi$ ambiguity. This would also be
consistent with the P.A. measured at 5 GHz. A detailed analysis is not
possible, since the images have too different angular resolutions, but
a rough comparison of the images leads to a rotation measure of about
185 rad/m$^2$. 

At 43.3 GHz we find a strong core and a weak jet. Unfortunately, the quality
of the data at this frequency was not good, and we were not able to
obtain a reliable calibration of the polarization data.

\section{Discussion}
\label{discussion}

\subsection{Spectrum}

The observations at 22 and 43 GHz add new pieces of information on
the spectrum of the jet in 3C\,264. In Table~\ref{tab2} we display the
data presented in Lara et al. (\cite{lara99}), adding new data from
our observations. The jet flux density at radio wavelengths has been
estimated in AIPS following these authors, that is by measuring the
total flux density and then subtracting the integrated flux density of
a Gaussian component that fits the core emission. In
Fig.~\ref{espec_jet}, we display the broad band spectrum of the jet
and determine a radio spectral index\footnote{The spectral index
$\alpha$ is defined so that the flux density $S \propto \nu^{\alpha}$}
$\alpha_r = -0.58$ from a least squares fitting to the data. The
spectrum between radio and optical wavelengths (spectral index of -1.34) is
consistent with a synchrotron emission mechanism and the existence of
a spectrum break frequency in the infrared. 

We also display in Table~\ref{tab2} and Fig.~\ref{espec_jet} the core
flux density determined at radio wavelengths (at 1.5, 5.0, 22.5 and
43.3 GHz) with an angular resolution of 90 mas. We obtain a flat
spectral index, with $\alpha_{core} = - 0.15$ (dashed line in
Fig.~\ref{espec_jet}).

\begin{figure}[]
\vspace{10cm}
\includegraphics{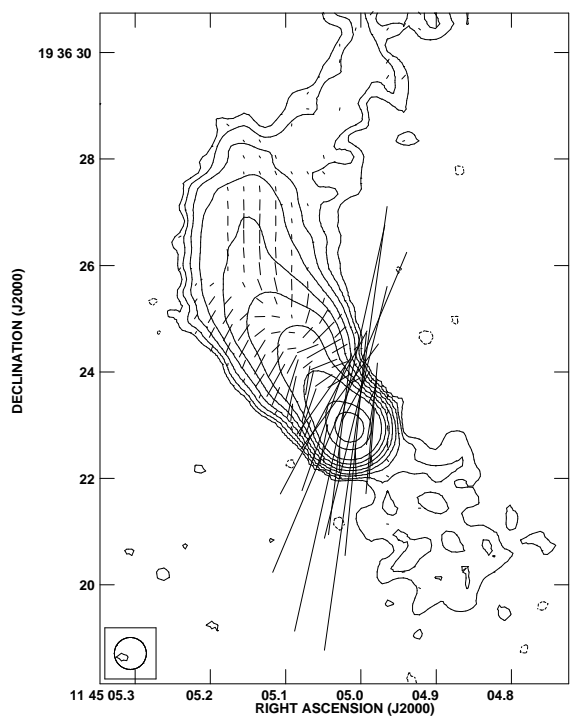} 
%%\rule{0.4pt}{4cm}% line thickness, height of picture
\caption{VLA map of 3C\,264 made at 5.0 GHz on 1999.68. We display
vectors representing the polarization position angle (E-vector), with
length proportional to the polarized flux ($1''\equiv 1.67$
mJy/beam). Contours are  
spaced by factors of 2 in brightness. The Gaussian beam used
for convolution is $0''.6 \times 0''.6$. The first contour level is 
0.13 mJy\,beam$^{-1}$ and the peak of brightness 0.225 Jy\,beam$^{-1}$.}
\label{bandac_polz}
\end{figure}

\begin{figure*}
\vspace{9cm}
\includegraphics{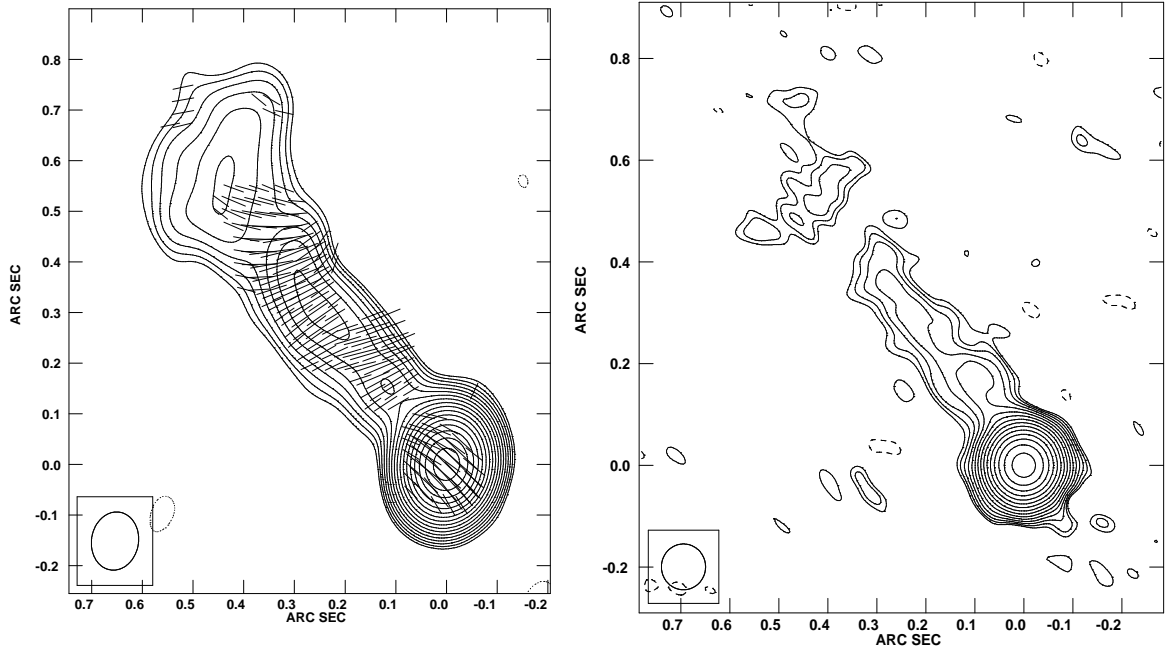}
%%\rule{0.4pt}{4cm}% line thickness, height of picture
\caption{VLA maps of 3C\,264 made at 22.5 (left) and 43.3 GHZ (right) on 
1999.68. At 22.5 GHz we display superimposed vectors representing the
polarization position angle (E-vector), with length proportional to
the polarized flux ($0''.1$ mas $\equiv$ 1 mJy\,beam$^{-1}$). In both
maps contours are spaced by factors of $\sqrt{2}$ in brightness. 
For both maps we list the Gaussian beam size used for convolution (in
mas), the first contour level 
(mJy\,beam$^{-1}$) and the peak of brightness (Jy\,beam$^{-1}$). {\bf
Left:} beam = 115 mas $\times$ 93 mas, P.\,A. $-9^{\circ}$; 1st
cntr. = 0.46; peak = 0.146 {\bf Right:} beam = 90 mas $\times$ 90 mas;
1st cntr. = 0.8; peak = 0.124.}
\label{bandakq}
\end{figure*}

\begin{table}[]
\caption[]{3C\,264 jet and core flux density}
\label{tab2}
\begin{tabular}{r@{.}l r@{.}l r@{.}l c}
\hline
\multicolumn{2}{c}{Frequency}  & \multicolumn{2}{c}{Jet Flux}  & \multicolumn{2}{c}{Core Flux} & References \\
  \multicolumn{2}{c}{(GHz)}    & \multicolumn{2}{c}{(mJy)}     & \multicolumn{2}{c}{(mJy)}     &            \\
\hline		       		   
  1&66      & 130&0     & 209&0     & 1 \\
  5&0       &  78&0     & 171&0     & 2 \\
 22&5       &  29&3     & 147&2     & 1 \\
 43&3       &  20&5     & 123&8     & 1 \\
  3&83\,e+5 &   0&076   & \mc       & 2 \\
  4&35\,e+5 &   0&062   & \mc       & 3 \\
  4&46\,e+5 &   0&049   & \mc       & 2 \\
  5&48\,e+5 &   0&046   & \mc       & 2 \\
 24&20\,e+5 &$<$0&0005  & \mc       & 4 \\ 
\hline
\multicolumn{7}{l}
{\parbox{7cm}{\footnotesize{References for the jet flux density: (1) These data; (2) Lara et al. \cite{lara99}; (3) Baum et al. \cite{baum2}; (4) Prieto et al. \cite{prieto}.}}}
\end{tabular}
\end{table}

\begin{figure}
\vspace{7cm}
\includegraphics{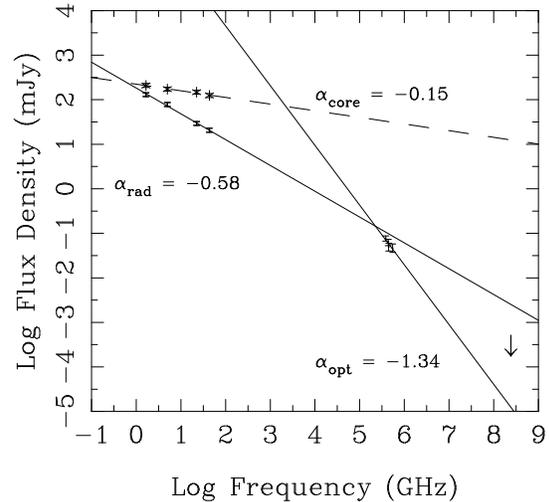}
%%\rule{0.4pt}{4cm}% line thickness, height of picture
\caption{Flux density of the jet and the core of 3C\,264 vs. frequency. 
Straight lines represent independent linear fits for the radio and
optical jet data. The dashed line represent the same for the core
radio data. Numerical values of the flux density are displayed in
Table~\ref{tab2}} 
\label{espec_jet}
\end{figure}

\subsection{Jet velocity and orientation}
\label{r}

Assuming that 3C\,264 consists of two intrinsically symmetric jets
with isotropic emissivity, their velocity, $\beta$, and orientation
with respect to the observer, $\theta$, can be constrained from the
jet to counterjet brightness ratio, $R$ (see Giovannini et
al. \cite{giov94} for more details). Values of $R$ at different
distances from the core have been compiled from the literature or
derived from our data considering the best available images at
different angular resolutions to assure that no jet brightness is
missing because of excessive angular resolution or poor sampling of the
visibility function. In Table~\ref{tab4} we report values (or limits)
for $R$, and the corresponding values (or limits) for
$\beta\cos\theta$ from radio and optical images, where we assume a
radio jet spectral index $\alpha = -0.5$ and an optical spectral index
$\alpha = -1.3$ (Lara et al. \cite{lara99}). The available data at
about $0''.5$ from the core constrain the angle $\theta$ to be $<
54^{\circ}$ with a velocity $\beta > 0.9$c.  Starting from $1''.5$
from the core we detect also the counterjet in the VLA images which
allows the estimation of values for $\beta$ and $\cos\theta$.

\begin{figure*}[]
\vspace{5cm}
\includegraphics{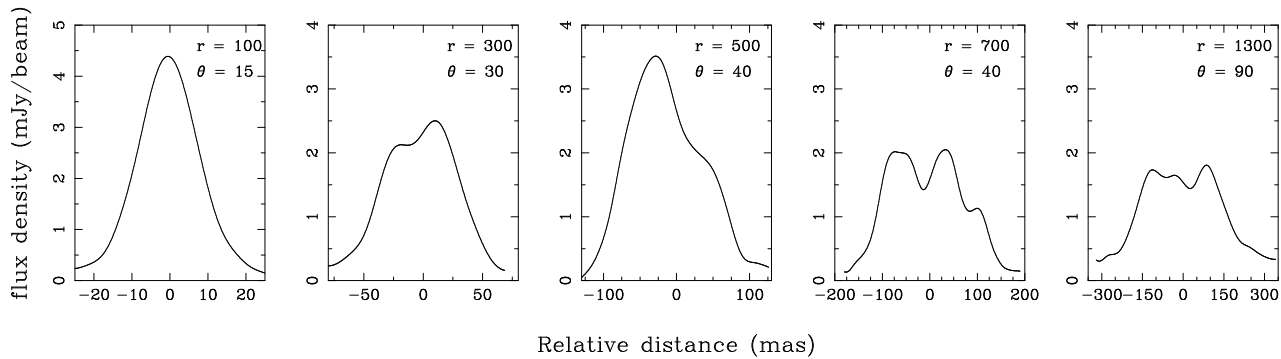}
%%\rule{0.4pt}{4cm}% line thickness, height of picture
\caption{Brightness profiles across the jet of 3C\,264. The distance
from the core (r) at which the profile is measured and the FWHM of the
circular Gaussian beam used for convolution ($\theta$), both in
milliarcseconds, are indicated on the upper right of each panel.}
\label{jetwidth}
\end{figure*}

The counterjet detection at 1$''$.5 and 2$''$ from the core constrains
the jet velocity to be $\sim 0.8$c at 600 pc from the core. Therefore,
the present results show that the decrease in the jet velocity is
slower than estimated in Baum et al. (\cite{baum2}). By comparing our
results with the adiabatic model discussed in Baum et
al. (\cite{baum2}) we find general agreement assuming an initial jet
velocity $\beta_{ini} = 0.99$ ($\gamma = 7$, where $\gamma=
1/\sqrt{1-\beta^2}$ is the Lorentz factor) and an orientation $\theta=
50^{\circ}$. A slower initial velocity predicts too low a jet velocity
in the range 400 to 600 pc from the core; a smaller value of $\theta$
is not possible for the same reason, while $\theta > 50^{\circ}$ is
not allowed from the jet to counterjet brightness ratio (see their
Figs. 18a-b).

\begin{table*}[]
\caption[]{Jet to counterjet brightness ratio}
\label{tab4}
\begin{tabular}{r@{.}l r@{.}l r lrrrc}
\hline
\multicolumn{4}{c}{Distance} & R & Instrument  &
$\beta\cos\theta$$^\ast$ &  $\theta_{max}$ & $\beta(50^{\circ})$  &
References\\
\multicolumn{2}{c}{($''$)} & \multicolumn{2}{c}{(pc)}&  &   &  &
($^{\circ}$) & &    \\ 
\hline
~0&005 &   2&1   & $>$24  & VLBI       & $>$0.56 & $<$56 & $>$0.87 & 1 \\
 0&01  &   4&1   & $>$18  & VLBI       & $>$0.52 & $<$59 & $>$0.81 & 1 \\
 0&05  &  20&6   & $>$23  & EVN+MERLIN & $>$0.55 & $<$56 & $>$0.86 & 2 \\
 0&10  &  41&2   & $>$16  & EVN+MERLIN & $>$0.50 & $<$60 & $>$0.78 & 2 \\
 0&15  &  61&8   & $>$20  & MERLIN     & $>$0.54 & $<$58 & $>$0.84 & 1 \\
 0&3   & 123&6   & $>$18  & MERLIN     & $>$0.52 & $<$59 & $>$0.81 & 1 \\
 0&3   & 123&6   & $>$37  & HST        & $>$0.49 & $<$60 & $>$0.77 & 1 \\
 0&5   & 206&0   & $>$28  & MERLIN     & $>$0.58 & $<$54 & $>$0.91 & 1 \\
 0&5   & 206&0   & $>$32  & HST        & $>$0.48 & $<$62 & $>$0.74 & 1 \\
 0&8   & 329&6   & $>$32  & MERLIN     & $>$0.60 & $<$53 & $>$0.93 & 2 \\
 1&0   & 412&0   & $>$16  & MERLIN     & $>$0.50 & $<$60 & $>$0.78 & 2 \\
 1&5   & 618&0   &    17  & VLA-A      &    0.51 &    59 &    0.80 & 2 \\
 2&0   & 824&0   &    15  & VLA-A      &    0.49 &    60 &    0.77 & 2 \\
 2&8   &1153&6   &    11  & VLA-A      &    0.45 &    64 &    0.69 & 2 \\
 3&3   &1359&6   &     5  & VLA-B      &    0.31 &    72 &    0.48 & 1 \\
 5&0   &2060&0   &     3  & VLA-B      &    0.22 &    78 &    0.34 & 1 \\
\hline
\multicolumn{10}{l}
{\parbox{12cm}{\footnotesize{$^\ast$ $\alpha = -0.5$ is assumed,
except with HST data where a valued of $\alpha = -1.3$ is assumed.\\ 
$\beta(50^{\circ})$ is the estimated jet velocity if $\theta = 50^{\circ}$\\
References: (1) Baum et al. \cite{baum2}; (2) Data in this paper}}}
\end{tabular}
\end{table*}

\subsection{Jet structure}

\subsubsection{Longitudinal structure}
\label{long}

The detailed images at 1.6 GHz of the first kiloparsec of the jet in
3C\,264 basically confirm the different regions described by Lara et
al. (\cite{lara99}) from observations at 5.0 GHz. A good
correspondence between features of the jet is observed at both
frequencies. The jet appears very well collimated up to a distance of
200 mas (82 pc) from the core. At this position, the jet undergoes a
strong widening, with its edges apparently defining a perfect cone
with an opening angle of $\sim 20^{\circ}$ (see
Fig.~\ref{combined}-right). This observed opening angle corresponds to
an intrinsic opening angle of $\sim 15^{\circ}$ (assuming an
orientation angle $\theta =
50^{\circ}$). The expected velocity for a freely expanding jet and an
intrinsic opening angle $\sim 15^{\circ}$ is $\beta = 0.96$
($\gamma=3.7$), in good agreement with values reported in Table~\ref{tab4}

Lara et al. (\cite{lara99}) find that the
brightness distribution of the jet at 5.0 GHz is not continuous
between 80 and 300 pc, showing three main ``blobs'' which might
suggest the existence of a helical twist. At 1.6 GHz, the observed
structure is also consistent with a possible helical filament embedded
in a conical nozzle. At $0''.8$ (330 pc) from the core we observe another 
change of properties: there seems to be a deflection of the
jet ridge line northward at this position, and the surface brightness
is significantly lower beyond. This distance coincides with the length
of the observed jet at 22.5 GHz. Moreover, as previously mentioned,
the electric vector at 1.6 GHz changes abruptly at $0''.8$ from the
core, from P.\,A. $\sim -40^{\circ}$ to P.\,A. $\sim -100^{\circ}$.
At further distances, the jet as observed at 1.6 GHz remains broad but
is again well collimated with a small opening angle.

\subsubsection{Transverse structure}

In Fig.~\ref{jetwidth} we present profiles of the surface brightness
distribution taken perpendicularly to the jet ridge line at several
positions along the jet. The profile at a distance of 100 mas from the
core is consistent with a spine brightened jet, with a FWHM of 20 mas
(broader than the FWHM of the Gaussian beam used for convolution of
the image). At smaller distances from the core, a spine brightened jet
is consistent with the VLBI image of 3C\,264 (Fig.~\ref{allmaps}). 
At distances from the core beyond 300 mas, the profiles become flatter
and are no longer indicative of spine brightening, being compatible
instead with an edge brightened jet. 

The apparent transition from a spine brightened to an edge brightened
jet is consistent with a two velocity jet regime. External jet regions
are decelerated by their interaction with the interstellar medium and
their velocity becomes significantly different from the jet spine
velocity. We note that with an orientation of $\sim 50^{\circ}$ with
respect to the observer, a high velocity jet is strongly de-boosted:
the Doppler factor $\delta = (\gamma (1-\beta \cos\theta))^-1$ is less
than 1, becoming smaller for increasing velocities. In this region
lower velocity structures are less de-boosted. Therefore the external
lower velocity region of the jet appears brighter than the inner high
velocity spine and we thus observe a limb-brightened jet structure. Of course,
at large distances from the core, where the inner spine is no more
strongly relativistic, the Doppler de-boosting is not relevant and
we observe again a jet with an uniform brightness distribution while in
the inner pc scale regions the jet can appear centrally peaked because
of angular resolution limitations or because the two velocity regime has
not started yet. We note that the existence of a velocity
structure in 3C\,264 is also consistent with Chiaberge et al.
(\cite{chiaberge}), who propose jet velocity structures as a general
scenario to reconcile the radio and optical properties of FR Is and BL
Lacs with the unification schemes.

The magnetic field observed
with MERLIN at 1.6 GHz (Fig.~\ref{allmaps}) and with the VLA at 22.5
GHz (Fig.~\ref{bandakq}-left) does not show the distribution expected by the 
models of jets advancing through the external medium, 
that is a perpendicular magnetic field
associated with the fast spine, and a predominantly parallel magnetic field
associated with the shear layer interacting with the external medium
(Laing et al. \cite{laing}; Aloy et al. \cite{aloy}). Our results
could be explained if most of 
the polarized emission in 3C\,264 comes from the shear layer of the
jet, as is effectively observed at larger distances at 5 GHz
(Fig.~\ref{bandac_polz}).  

\subsection{Disk or bubble at the center of NGC\,3862?}

\begin{figure}
\vspace{5.5cm}
\includegraphics{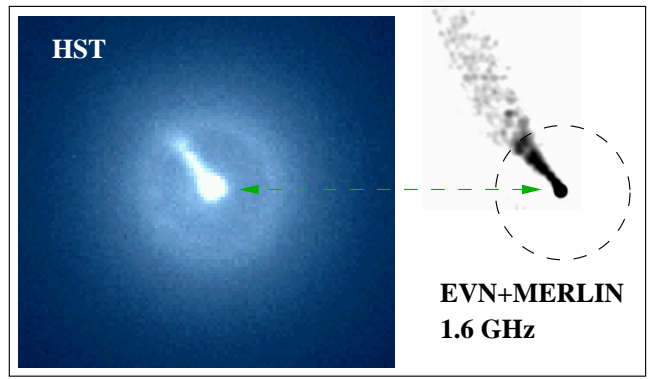}
%%\rule{0.4pt}{4cm}% line thickness, height of picture
\caption{{\bf Left:} HST archive image of 3C\,264 showing the
circumnuclear ring and the optical jet (Baum et
al. \cite{baum2}). {\bf Right:} Radio image of the jet at 1.6 GHz,
with an angular resolution of 40 mas. The locus of the ring seen in
the optical is plotted on the radio image for comparison.}
\label{hstradio}
\end{figure}

The position where the second important change of properties in the
jet of 3C\,264 takes place, that is at $0''.8$ from the core, is
precisely coincident with the radius of the optical ring, as already
noted by Baum et al. (\cite{baum2}; see Fig.~\ref{hstradio}). 
It is not clear from optical
studies of the host galaxy of 3C\,264, NGC\,3862, if this ring defines
the boundaries of a face-on flattened disk, or rather of a central
bubble. Disk structures are commonly found in the central kpc regions
of the host galaxies of FR I radio galaxies (e.g. de Koff et
al. \cite{dekoff}, Martel et al. \cite{martel}), strongly supporting
the face-on disk scenario in NGC\,3862. However, Hutchings et
al. (\cite{hutchings}) suggest that the ring defines an evacuated
region cleared of dust by some nuclear-related process.  The
analysis of the color excess and gradients inside the ring of
NGC\,3862 by Martel et al.  (\cite{martel}) support the existence ``of
a partially evacuated, inclined, and flattened system surrounded by a
red ring''.

From the current optical and radio data it is not possible to
discriminate between the face-on disk or the spherical bubble
scenario. If we consider the possibility of a face-on disk, the change in
the jet properties at the position of the optical ring would be just a
coincidence observed in projection on the disk boundary (see
Fig.~\ref{hstradio}). Moreover, that would also imply a closer than
inferred orientation of the jet toward the observer, assuming that
disks and jets are approximately perpendicular (e.g. Capetti \&
Celotti \cite{capetti1}). In this context, we note also the claim by
Sparks et al. (\cite{sparks}), suggesting that optical synchrotron
emission is observed only when the jet points toward the observer,
based on the fact that 4 out of 5 nearby FR I radio galaxies with
optical synchrotron jets present almost circular face-on dust
disks. This suggestion assumes a face-on disk scenario and jets
perpendicular to the disks. However, they explicitly mention 3C\,264
as a peculiar case.

On the other hand, if we consider the possibility of an evacuated spherical
bubble in the center of NGC\,3862, it would render a natural
explanation for the almost perfect circular symmetry of the ring and
the change of the radio (and optical) properties of the jet as it
crosses the bubble boundary (Baum et al. \cite{baum2}; Hutchings et
al. \cite{hutchings}).  Moreover, the observed widening of the jet
inside the ring would be consistent with an expansion favored by the
lower density medium within the evacuated bubble. This scenario would
also be consistent with an orientation of the jet of 3C\,264 of the
order of $50^{\circ}$ with respect to the observer (Sect. \ref{r}).

\section{Conclusions}

We present in this paper new observations of the radio source 3C\,264
which are sensitive to linear scales ranging from the parsec to
several kiloparsecs. Observations at 1.6 GHz were made simultaneously
with the EVN and MERLIN in 1997. Observations at 1.6, 5.0, 22.5 and
43.3 GHz were made two years later with the VLA in its most extended 
configuration. The main conclusions are:

\begin{enumerate}
\item The observations confirm the existence of different regions in the jet
of 3C\,264, as previously found by Lara et al. (\cite{lara99}). The
most remarkable feature is the transition between a well collimated
narrow jet at distances from the core below 80 pc, to a conical shaped
wide jet, with an opening angle of $20^{\circ}$. This transition
coincides also with a transition in the transverse structure of the
jet, which is consistent with a spine brightened jet in the collimated
region and an edge brightened jet in the conical region.  This result
is also in agreement with the possible presence of jet velocity structure
(fast spine surrounded by a slower layer) as found in other FR I
sources (e.g. \object{B2\,1144+35}; Giovannini et al. \cite{giov99})
or Bl Lacs (Mkn\,501; Giroletti et al. \cite{giroletti})

\item An apparent
deflection of the jet ridge line, together with a diminution of the surface
brightness and a change in the polarization properties, is present at
a distance of $\sim 300$ pc from the core, strikingly coincident with
the radius of a ring observed at optical wavelengths.

\item From observational data we estimate an initial
jet velocity $\beta=0.99$ and an orientation of the jet with respect
to the observer $\theta = 50^{\circ}$, consistent with previous
estimations of these parameters (Baum et al. \cite{baum2}).

\item Our data contribute to a better knowledge of the spectrum of the
radio-optical jet of 3C\,264. We find a radio spectral index of -0.58
which has to be compared with a spectral index of -1.34 at optical
wavelengths. The spectrum between radio and optical wavelengths is
consistent with a synchrotron emission mechanism and the existence of
a spectrum break frequency in the infrared. 

\item The properties of the jet structure, the changes in
the polarization angle, and the plausible jet orientation ($\sim
50^{\circ}$) with respect to the observer can be naturally brought
into agreement with the presence of a spherical bubble in the center
of NGC\,3862, even if the existence of a face on dust and gas disk
cannot be completely ruled out. 
\end{enumerate}

\begin{acknowledgements}

We thank Dr. Muxlow for his assistance during the MERLIN data
reduction process at Jodrell Bank; L.L. acknowledges support by the
European Union under contract ERBFMGECT950012. This research is
supported in part by the Spanish DGICYT (AYA2001-2147-C02-01). LF and GG
acknowledges the Italian Ministry for University and Research (MURST)
for financial support under grant Cofin 2001-02-8773. The MERLIN/VLBI
National Facility is operated by the University of Manchester on
behalf of the Particle Physics and Astronomy Research Council. The
European VLBI Network is a joint facility of European, Chinese and
other radio astronomy institutes funded by their national research
councils.  The National Radio Astronomy Observatory is a facility of
the National Science Foundation operated under cooperative agreement
by Associated Universities, Inc.

\end{acknowledgements}


\begin{thebibliography}{Normandin \& Kronberg 1980}

\bibitem[2000]{aloy} Aloy, M.A., G\'{o}mez, J.L., Ib\'{a}\~nez, J.M., et al. 2000, \apj, 582, L88

\bibitem[1988]{baum1} Baum, S. A., Heckman, T., Bridle, A., et al. 1988, \apjs, 68, 643

\bibitem[1997]{baum2} Baum, S. A., O'Dea, C. P., Giovannini, G., et al. 1997, \apj, 483, 178 (Erratum 492, 854 (1998))

\bibitem[1981]{bridle} Bridle, A. H. \& Vall\'ee, J. P. 1981, \anj, 86, 1165 

\bibitem[1999]{capetti1} Capetti, A. \& Celotti, A. 1999, \mn, 304, 434 

\bibitem[2000]{capetti2} Capetti, A., Trussoni, E., Celotti, A. et
al. 2000, \mn, 318, 493

\bibitem[2000]{chiaberge} Chiaberge, M., Celotti, A., Capetti, A. \& Ghisellini, G. 2000, \aa, 358, 104

\bibitem[1993]{crane} Crane, P., Peletier, R., Baxter, D. et al. 1993, \apj, 402, L37 

\bibitem[1995]{dekoff} De Koff, S., Baum, S., Sparks, W. B. et al. 1995, \baas, 187, 1202

\bibitem[1974]{fanaroff} Fanaroff, B. L. \& Riley, J. M. 1974, \mn, 167, 31

\bibitem[1981]{gavazzi} Gavazzi, G., Perola, C. G. \& Jaffe, W. 1981, \aa, 103, 35 

\bibitem[1994]{giov94} Giovannini, G., Feretti, L., Venturi, T. et al. 1994, \apj, 435, 116

\bibitem[1999]{giov99} Giovannini, G., Taylor, G. B., Arbizzani, E. et al. 1999, \apj, 522, 101

\bibitem[2001]{giov01} Giovannini, G., Cotton, W. D., Feretti, L., et al. 2001, \apj, 552, 508 

\bibitem[2003]{giroletti} Giroletti, M., Giovannini, G., Feretti, L.,
et al., \apj, in press

\bibitem[1998]{hutchings} Hutchings, J. B., Baum, S. A., Weistrop, D., et al. 1998, \anj, 116, 634

\bibitem[1993]{laing} Laing, R. A. 1993, in Space Telescope Sci. Inst. Symp. 6: Astrophysical Jets, ed. D. Burgarella, M. Livio \& C. P. O'Dea, Cambridge University Press, Cambridge, 95

\bibitem[1997]{lara97} Lara, L., Cotton, W. D., Feretti, L., et al. 1997, \apj, 474, 179

\bibitem[1999]{lara99} Lara, L., Feretti, L., Giovannini, G., et al. 1999, \apj, 513, 197

\bibitem[2000]{martel} Martel, A. R., Turner, N. J., Sparks, W. B., Baum, S. A. 2000, \apjs, 130, 267 

\bibitem[1988]{muxlow} Muxlow, T. W. B., Junor, W., Spencer, R. E., et al. 1988, in IAU Symp. 129: The Impact of VLBI on Astrophysics and Geophysics, ed. M. J. Reid M. J. \& J. M. Moran, Kluwer Academic Publishers, 131

\bibitem[1996]{prieto} Prieto, M. A. 1996, \mn, 282, 421

\bibitem[1994]{shepherd} Shepherd, M. C., Pearson, T. J., \& Taylor, G. B. 1994, \baas, 26, 987

\bibitem[2000]{smith} Smith, R. J., Lucey, J. R., Hudson, M. J., et al. 2000, \mn, 313, 469

\bibitem[2000]{sparks} Sparks, W. B., Baum, S. A., Biretta, J. et al. 2000, \apj, 542, 667

\end{thebibliography}
\end{document}